\documentstyle[12pt]{article}
\begin{document}

\title{Classicality Criteria}
\author{
Nuno Costa Dias\footnote{{\it ncdias@mail.telepac.pt}} \\
{\it Departamento de Matem\'{a}tica} \\
{\it
Universidade Lus\'ofona de Humanidades e Tecnologias} \\  {\it Av. Campo Grande, 376,
1749-024 Lisboa, Portugal}}
\maketitle

\begin{abstract}
We present two possible criteria quantifying the degree of classicality of an arbitrary (finite dimensional) dynamical system. The inputs for these criteria are the classical dynamical structure of the system together with the quantum and the classical data providing the two alternative descriptions of its initial time configuration. It is proved that a general quantum system satisfying the criteria up to some extend displays a time evolution consistent with the classical predictions up to some degree and thus it is argued that the criteria provide a suitable measure of classicality.
The features of the formalism are illustrated through two simple examples.
\end{abstract}

\section{Introduction}

It is generally accepted that quantum mechanics yields the most fundamental description of all physical systems. Such status requires the theory to provide a suitable description of every day classical like phenomena. However, quantum mechanics faces a considerable number of problems to explain the emergence of a classical domain.
In a broad sense, this is called the problem of the semiclassical limit of quantum mechanics \cite{neumann,wheeler,hartle1,halliwell1,halliwell2,robert1}.

This is a difficult topic not least because there are a variety of
different ways of relating classical and quantum mechanics. The
standard approach is to take the limit $\hbar \to 0$, or
equivalently, the limit of a large number of particles $N \to
\infty$. Under some general conditions one can derive the
classical evolution from the quantum dynamics \cite{paul1}.
Alternatively, one may start with a set of quantum initial data
with minimal spread and show that its quantum time evolution
remains very peaked around the classical orbits. Coherent states
\cite{glauber} (and squeezed states \cite{combescure2,hagedorn1})
have been extensively studied in this context
\cite{hepp,heller1,heller2,combescure3,combescure4,hagedorn2,hagedorn3}.
It has been proved, using several different approaches and for a
large class of dynamical systems, that coherent states evolve, up
to a prescribed error and during a prescribed lapse of time,
peaked around the classical paths
\cite{combescure3,hagedorn2,hagedorn3,bievre}. Coherent states
methods have also been used to prove that in the limit $\hbar \to
0$ and for times shorter than Eherenfest's time, the quantum and
the classical predictions coincide exactly
\cite{hagedorn1,bievre,bievre2,bievre3}. Related results have been
obtained using the tools of pseudodifferential calculus
\cite{hormander,sjostrand,yajima} and microlocal analysis
\cite{helffer1,helffer2,robert2}. These have played an important
part in several attempts to understand the semiclassical behavior
of classically chaotic systems \cite{helffer1,zelditch}.

Two other significant approaches are the Wigner (or more generally
the quasidistributions) \cite{lee} and the dechoerent histories
\cite{hartle1,halliwell1,hartle2} approaches. The aim is to
(either by identifying proper quantum phase space distributions,
or by coupling the quantum system to a suitable environment)
recover the classical statistical predictions from the quantum
formulation of the system. All these views are of course not
antagonic, but complementary, meaning that the classical behavior
can emerge in a variety of different regimes.

Despite the importance of these results there are still several open problems in the field of the semiclassical limit of quantum mechanics.
The common view is that classical mechanics is an approximate description of quantum mechanics. This approximation is not always valid. In fact, one expects it to be only valid for some particular set of quantum initial data and even in this case not for all dynamical systems (consider for instance a particle hitting a potential barrier).
An important question is then what is the degree of validity of the classical approximation? Clearly this question does not have a single answer. It depends on the dynamical structure of the system and on the quantum and also the classical data providing the two alternative descriptions of the initial time configuration.
Let us assume that we are given this information. Can we compare the classical and the quantum descriptions of the dynamical system and produce a statement about the degree of validity of the classical description? Or, in other words, can we produce a statement about the degree of classicality of the quantum description?
The answer to these questions should be affirmative since we are given all the necessary ingredients. However, and despite the fact that the pertinence of these questions have been recognized by several authors \cite{hartle1,halliwell1,halliwell2}, the problem of establishing a measure of classicality is still lacking an unique solution.

Closely related to this topic is the problem of establishing a clear relation between the quantum initial state and the classical like behavior. Very few studies elaborate on this subject. Instead, one typically assumes
a particular quantum initial state and works from there to prove the "classicality" of the quantum dynamics. A weak point in the procedure is that we are still lacking a precise definition of "classicality".
For instance in \cite{hagedorn1,combescure3,hagedorn2} the aim is to provide minimal bounds for the spreading of the time evolution of a specific quantum initial data (typically a coherent or squeezed state). But the questions that remain are: are these results extendable to a wider class of quantum initial states, or more generally, what is the relation between the spreading of the wave function through time and the quantum initial state? And further, how is the spreading of the wave function related to classicality?

In this paper we attempt to provide a partial answer to this questions. In general terms we study the inverse problem of \cite{hagedorn1,combescure3,hagedorn2}: we start by imposing the bounds (which will be related to the error margins of the classical time evolution) and attempt to identify those initial quantum states which display a time evolution that satisfies these bounds for all times.
In doing so we identify a (possible) set of general conditions an arbitrary, finite dimensional quantum system should satisfy so that it evolves in agreement with the classical predictions. These conditions will be of the form of a sequence of restrictions on the initial quantum state (the higher the classicality the more stringent the conditions) and provide the basic mathematical structure to construct a measure of classicality.

More precisely our approach will be as follows:
We work in the context of the standard {\it Copenhagen} formulation of quantum mechanics \cite{heisenberg,bohr,dirac1,cohen} and
consider a general dynamical system with $N$ degrees of freedom. Its initial time configuration might be described by a set of quantum initial data or, alternatively, by a set of classical initial data.
The classical description is given by a set of values
$a_i^0$ for a complete set of classical observables $a_i$ ($i=1...2N$), together with their associated error margins $\delta _i$. The quantum description is given by the initial time wave function
$|\psi >$. Classical mechanics states that the output of a measurement of an observable
$a_i$ will belong to the interval $[a_i^0-\delta_i,a_i^0+\delta_i]$.
Quantum mechanics, in turn, states that there is some probability $p_i \le 1$ that a measurement of the observable $\hat{a}_i$ yields a value inside the former interval. Clearly the two statements are not equivalent. However, there are a number of fairly intuitive ways by which one can measure the consistency (i.e. the agreement) of the two former descriptions of the initial time configuration. We will propose two such {\it consistency criteria} in section 2.

Unfortunately, the fact that the two descriptions of the initial time configuration are consistent up to some degree does not imply that the classical and the quantum descriptions of a future configuration will also be consistent up to the same degree. Our aim is then to identify the conditions that should be satisfied by the initial time configuration
so that the degree of consistency is preserved through time evolution.
These conditions are obtained (they are given by a set of relations between the classical initial data ($a_i^0,\delta_i$) and the spreads of the initial wave function $|\psi>$) and used to construct two alternative classicality criteria.
Given the classical initial data these conditions constitute a sequence of growing restrictions on the functional form of the initial data wave function.
When $|\psi>$ satisfies the $n$ first former conditions we say that $|\psi>$ is $n$-{\it order classical}.
When $|\psi>$ satisfies the full set of conditions we say that $|\psi>$
is a {\it classical limit initial data wave function}.

The main property of the classicality criteria is that if the classical and the quantum initial data satisfy the classicality criteria up to some degree then the time evolution of the classical initial data (obtained using classical mechanics) and the time evolution of the quantum initial data (obtained using quantum mechanics) will display a minimum degree of consistency for all times. Hence, the classicality criteria supply a suitable (although clearly not unique) measure of classicality that might be computed, for an arbitrary dynamical system, at the kinematical level.

These results are not limited to some specific type of systems displaying a particular dynamical behavior or having some particular quantum initial data. On the contrary, for an arbitrary dynamical system the classicality criteria can be used to determine the set of quantum initial data that displays a $n$-order classical time evolution (a quantum time evolution $n$-order consistent with the classical predictions). This will be done explicitly for the two physical examples presented in sections 6 and 7.

Further applications of the criteria are given in \cite{nuno2,nosso1,nosso2}.

\section{Consistency criteria}

Let us consider a general dynamical system with $N$ degrees of freedom. The phase space $T^{\ast}M$ has the structure of the cotangent bundle of the configuration space $M$, and  henceforth will be also assumed to have the structure of a {\it flat} sympletic manifold.
Therefore a global Darboux chart can be defined in $T^{\ast}M$. Let us choose a set of canonical variables $a_i, i=1..2N$ (where $a_i=q_i, i=1..N$ and $a_i=p_{i-N}, i=N+1..2N$). They yield the sympletic structure as the 2-form $w=dq_i \wedge dp_i$.

The classical description of a specific time configuration of the system is given by a set of values $a_i^0$ for the complete set of observables $a_i$ together with the associated error margins $\delta_i$. The classical statement is that a measurement of $a_i$ will yield a value $a_i^1$ belonging to the classical error interval $I_i = [a_i^0 - \delta _i, a_i^0 + \delta_i]$.

Alternatively, quantum mechanics describes the same configuration of the system with a wave function $|\psi>$ belonging to the physical Hilbert space ${\cal H}$ and the quantum statement is that there is some probability
$p(a_i)=\sum_{k}|<a_i,k|\psi>|^2$ that a measurement of the observable $\hat{a}_i$ yields the value $a_i$, (where the states $|a_i,k>$ form a complete set of eigenvectores of $\hat{a}_i$ that spans the Hilbert space ${\cal H}$, $a_i$ are the associated eigenvalues and $k$ is the degeneracy index). For notational convenience we will assume that $\hat A$ displays a discrete spectrum, but all results can be easily rewritten for the continuous case.

A straightforward way of measuring the consistency between the two descriptions is given by the following criterion:

\underline{{\bf Definition}} {\bf - First consistency criterion}\\
Let us calculate the probabilities $p_i$ - generated by the wave function $|\psi>$ in the representation of each of the observables $\hat{a}_i$, $i=1..2N$ - that a measurement of $\hat a_i$ yields a value belonging to the classical error interval $I_i=[a_i^0 - \delta _i, a_i^0 + \delta_i]$, i.e.:
\begin{equation}
p_i= \sum_{a_i \in I_i , k} |<a_i,k|\psi>|^2.
\end{equation}
The minimum value of the set $\{ p_i: i=1..2N\}$ provides a suitable measure of the consistency between the classical and the quantum descriptions of the configuration of the dynamical system. If this minimum value above is for instance $p_0$ we say that the classical and the quantum descriptions are $p_0$-{\it consistent}. Furthermore, if $p_0=1$ then the two descriptions are fully consistent and $|\psi>$ is named a {\it classical limit wave function}.$_{\Box}$

Another, probably less intuitive, consistency criterion is given by the following definition:

\underline{{\bf Definition}} {\bf - Second consistency criterion}\\
Let $0 \le p < 1$ be an arbitrary probability and let $M\in {\cal N}$. To each classical observable $a_i$ we associate the set of intervals of the form:
\begin{equation}
I_i(p,M)=[a_i^0 - \frac{\delta_i}{(1-p)^{1/(2M)}},a_i^0 +
\frac{\delta_i}{(1-p)^{1/(2M)}}] .
\end{equation}
In each of the former intervals we can calculate the probability $p_i$ generated by the
wave function $|\psi>$, in the representation of the corresponding quantum observable $\hat{a}_i$:
\begin{equation}
p_i(p,M)= \sum_{a_i \in I_i(p,M), k} |<a_i,k|\psi>|^2.
\end{equation}
For given values of the classical and quantum data $a_i^0$, $\delta_i$ and $|\psi>$ this probability is an exclusive function of $p$ and $M$.
The second consistency criterion is defined as follows:
The classical and the quantum data, describing a given configuration of the dynamical system, are $M$-{\it order consistent} if and only if for all $0 \le p <1$ and for all
$i=1..2N$ the condition $p_i(p,M) \ge p$ is satisfied, i.e.:
\begin{equation}
\sum_{a_i \in I_i(p,M), k} |<a_i,k|\psi>|^2 \ge p \quad ,
\forall_{p \in [0,1[}, \forall_{i=1..2N},
\end{equation}
where $I_i(p,M)$ is given by (2).
If $M=\infty$ the classical and the quantum descriptions are fully consistent, in which case $|\psi>$ is a {\it classical limit} wave function.$_{\Box}$

The former criterion provides a measure of how peaked is the wave function - in the representation of each of the quantum observables - around the classical error margin of the corresponding classical observable. This will became clear in the sequel (16).

As a first remark let us point out that the two alternative definitions of a classical limit wave function (using the first and the second consistency criterion) are, in fact, equivalent, i.e. $p=1 \Longleftrightarrow M=+\infty$, which in turn, is equivalent to the statement that:
\begin{equation}
|\psi>=\sum_{a_{i} \in I_i,k}  <a_{i},k|\psi>|a_{i},k>\quad , \forall_{i=1..2N},
\end{equation}
where, as before $I_i=[a^0_{i}-\delta_i,a^0_{i}+\delta_i]$, $(a_i^0,\delta_i)$ is the classical data and
$|a_{i},k>$ is the general eigenstate of the operator $\hat{a}_i$.

As a second remark let us notice that in some cases it is not possible to construct a wave function satisfying eq.(5) exactly. One should keep in mind that eq.(5) stands for the limit $M \to \infty$ or $p \to 1$ which may, or may not admit an explicit realization in terms of some wave function $|\psi>$.

Some emphasis will be put in the study of the time evolution of a general classical limit initial data wave function.
Although, for some systems, we might be unable to write $|\psi>$ (satisfying (5)) explicitly, its time evolution can be obtained, in the Heisenberg picture, using the standard rules of quantum mechanics.
This study will prove to be useful allowing us to develop a set of techniques that will later be used to obtain the classicality criteria and, in the sequel, to study physical systems with general initial data.

\section{Error ket framework}

Let us introduce dynamics by specifying a general Hamiltonian $H$.
The quantum time evolution of an arbitrary fundamental observable $\hat{A}$ (with $\hat{A}=\hat{a}_i$ for some $i=1..2N$) is given by:
\begin{equation}
\hat{A}(t)=\sum_{n=0}^{\infty}\frac{1}{n!}
 \left( {\frac{t}{i\hbar}} \right)^n [...[[\hat{A},\hat{H}],\hat{H}]...].
\end{equation}
Alternatively, the standard classical treatment of the same system provides the predictions:
\begin{equation}
{A}(t)=\sum_{n=0}^{\infty}\frac{t^n}{n!}
 \{...\{ \{ {A},{H} \},{H}\}...\} \qquad \qquad
\delta_A(t) = \sum_{n=1}^{+\infty} \frac{1}{n!} \sum_{k_1,...,k_n=1}^{2N} \left| \frac{\partial^n A(t)}{\partial a_{k_1}...\partial a_{k_n}} \right|
\delta_{k_1}...\delta_{k_n},
\end{equation}
where the error margins $\delta_{k_i}$ are taken at the initial time. Notice that, in general, it is possible to obtain more accurate predictions than those of (7), i.e. with smaller error margins. Such estimates have the disadvantage of requiring a case by case evaluation, while eq.(7) is valid in general.

To avoid some possible misunderstandings we will, from now on, focus on the case of the second consistency criterion. In section 5 our results will then be re-written using the language of the first consistency criterion. Let us then assume that the classical and the quantum initial data are $M$-consistent.
The question we would like to answer is then: which are the conditions that should be satisfied so that the quantum description of the system at a time $t$ - given by the initial data wave function $|\psi>$ in the representation of the observables $\hat{a}_i(t)$ (6) - is also $M$-consistent with the classical description given by (7)?

This is not an easy task. We start by presenting a framework that will later be used to give a precise answer to this question.

\subsection{Definition and properties of $|E>$ and $\Delta$}

Let us start by introducing the relevant definitions. Let $\hat{A}$ be an operator acting on the quantum Hilbert space ${\cal H}$.
Let $|a,k>$ be a complete set of eigenvectors of $\hat{A}$, with associated eigenvalues $a$ and $k$ being the degeneracy index. This set forms a complete orthogonal basis of ${\cal H}$. Finally, let $|\psi>$ be the wave function describing the system.

\underline{{\bf Definition}} {\bf - Error Ket}\\
We define the {\it nth-order error ket} $|E^n(\hat{A},\psi,a^0)>$,
as the quantity:
\begin{equation}
|E^n(\hat{A},\psi,a^0)>=(\hat{A} - a^0)^n |\psi>,
\end{equation}
where $n\in {\cal N}$, $a^0\in {\cal C}$ and $\hat{A}$ does not need to be self-adjoint.
The error bra $<E^n(\hat{A},\psi,a^0)|$ is defined accordingly to the definition of the error ket.

Let now $\hat{A_{z}}$, $z=1,..n$ be a set of operators acting on ${\cal H}$. For each value of $z$ let $|a_{z},k_z>$ be a complete set of eigenvectors of $\hat{A_z}$, with eigenvalues $a_{z}$, and $k_z$ being the degeneracy index. Moreover let $a_z^0 \in {\cal C}$.
The {\it nth-order mixed error ket} $|E(\hat{A}_{1},\hat{A}_{2},....\hat{A}_{n},\psi,
a_1^0,a_2^0,...,a_n^0)>$, is defined as the quantity:
\begin{equation}
|E(\hat{A}_{1},\hat{A}_{2},....\hat{A}_{n},\psi,a_1^0,a_2^0,...,a_n^0)>=
(\hat{A}_n - a_n^0).... (\hat{A}_2 - a_2^0) (\hat{A_1} - a_1^0) |\psi>.
\end{equation}
When there is no risk of confusion we will use the short notations $|E^n>$
or $|E^n_A>$ for the $n$th-order error ket and $|E_{A_1,A_2,...A_n}>$ for
the mixed error ket.$_{\Box}$

We shall now study some of the properties of this quantity:\\
{\bf a)} {\it Explicit form of the error ket}.\\
Let us start with the 1st-order error ket. We have: $|E(\hat{A},\psi,a^0)>=\sum_{a,k}(a-a^0)<a,k|\psi>|a,k> $.
This result is easily extended to the case of the $n$th-order
mixed error ket $|E_{A_1,A_2,...A_n}>$:
\begin{eqnarray}
&& |E_{A_1,A_2,...A_n}> =\sum_{a_1,k_1}\sum_{a_2,k_2}...\sum_{a_n,k_n}
(a_1-a^0_1) (a_2-a^0_2)....(a_n-a^0_n)
<a_1,k_1|\psi> \nonumber \\
&& <a_2,k_2|a_1,k_1><a_3,k_3|a_2,k_2>....
<a_n,k_n|a_{n-1},k_{n-1}>|a_n,k_n>,
\end{eqnarray}
and if $\hat{A}_1=
\hat{A}_2=....=\hat{A}_n=\hat{A}$ we have $|E_{A_1,A_2,...A_n}>=|E^n_{A}>$ which
give us the explicit expression for the $n$th-order error:
\begin{equation}
|E^n_{A}>=(\hat{A}-a^0)^n|\psi>
=\sum_{a,k}(a-a^0)^n <a,k|\psi>|a,k>.
\end{equation}
{\bf b)} {\it Relation between the nth-order error ket and the mean 2nd-order
deviation.} \\
Let us calculate the value of $<E_A^n|E_A^n>$:
\begin{equation}
<E_A^n|E_A^n>=\sum_{a,k} \left((a-a^0)^{\ast}(a-a^0)\right)^n
|<a,k|\psi>|^2,
\end{equation}
and if $\hat{A}$ is self-adjoint and $a^0=<\psi|\hat{A}|\psi>$
then $<E_A^n|E_A^n>$ is just the mean $2n$th-order deviation of $\hat{A}$:
$(\Delta\hat{A})^{2n}$
(if $n=1$ then $<E^1_A|E^1_A>$ is just the mean square deviation). \\
{\bf c)} {\it Nth-order error ket of a classical limit initial data wave function}.\\
Let us consider the case in which the initial time configuration of the dynamical system is described by a classical limit initial data wave function $|\psi>$. Let $A$ be a fundamental observable of the system and let the classical initial data associated to $A$
be given by $a^0$ with error margin $\delta_A$.
From eqs.(5) and (12) it is straightforward to obtain the relation:
\begin{equation}
<E_A^n|E_A^n>=\sum_{a,k} |a-a^0|^{2n}|<a,k|\psi>|^2
\le \delta_A^{2n} \sum_{a,k}|<a,k|\psi>|^2=\delta_A^{2n},
\end{equation}
which is valid for all $n \in {\cal N}$ and for all fundamental observables $A=a_i (i=1..2N)$.\\
{\bf d)} {\it Probabilistic distribution function from the nth-order error ket}.\\
Let $\hat{A}$ be self-adjoint, let $|\psi>$ be the state of the system and let $a^0$ be a real number. Given $<E^n(\hat{A},\psi,a^0)|E^n(\hat{A},\psi,a^0)>$, to each "quantity of probability" $0 \le p < 1$ we can associate an interval $I_n$ around $a^0$, $I_n=[a^0-\Delta_n,a^0+\Delta_n]$, such that the probability of obtaining a value $a \in I_n$ from a measurement of $\hat{A}$ is at least $p$. The range of the interval $I_n$ is dependent of $\hat{A} , \psi$ and $a^0$ only through the value of $<E_A^n|E_A^n>$. The quantity $\Delta_n(<E_A^n|E_A^n>,p)$ is named the {\it nth-order spread} of the wave function. Let us show that if $\Delta_n=\Delta_n(\hat{A},\psi,a^0,p)$ is given by:
\begin{equation}
\Delta_n(\hat{A},\psi,a^0,p)=\left(\frac{<E^n_A|E^n_A>}{1-p}\right)^{1/2n}
\end{equation}
then the probability of obtaining a value $a\in I_n$ from
a measurement of $\hat{A}$ is at least $p$. From (12,14) we have:
\begin{eqnarray}
(\Delta_n)^{2n}(1-p) & = & \sum_{k,a} |a-a^0|^{2n}
|<a,k|\psi>|^2 \nonumber \\
\ge \sum_{k,a\notin I_n} |a-a^0|^{2n}|<a,k|\psi>|^2
& \ge & (\Delta_n)^{2n} \sum_{k,a\notin I_n} |<a,k|\psi>|^2,
\end{eqnarray}
and this implies: $\sum_{k,a\notin I_n}|<a,k|\psi>|^2 \le 1-p $,
which is the result we were looking for. If $\hat{A}$ is not self-adjoint
the former result can also be obtained, but in this case $I_n$ is a ball of radius $\Delta_n$ in the complex plane.\\
{\bf e)} {\it Corollary of d)}.\\
A straightforward consequence of the previous result is the following: If for some positive integer $M$ the condition:
\begin{equation}
<E^M(\hat{a}_i,\psi,a^0_i)|E^M(\hat{a}_i,\psi,a_i^0)> \le \delta_i^{2M}
\end{equation}
holds for all $i=1..2N$ then the classical and the quantum descriptions (given by ($a_i^0, \delta_i, i=1...2N$) and $|\psi>$, respectively)
are, accordingly to the second consistency criterion, $M$-order consistent.

\subsection{Time evolution of $|E>$ and $\Delta$}

The aim is now to determine the $n$th-order error ket and the $n$th-order spread associated to the operator $\hat{A}(t)$ given in (6), as a function of the error kets and spreads associated with the initial time operators. The calculations might seem, in a first reading, complicated and cumbersome. Nevertheless, the final result will be simple and physically appealing.

\subsubsection{Evolving the error Ket}

We will start by calculating the 1st-order error ket associated with the sum and with the product of two arbitrary operators $\hat{A}$ and $\hat{B}$ and with the product of an operator by a scalar. Let the state of the system be $|\psi>$, let $|E(\hat{A},\psi,a)>$ and $|E(\hat{B},\psi,b>$ be the 1st-order error kets associated to $\hat{A}$ and $\hat{B}$ and let $a,b,c \in {\cal C}$.

\underline{{\bf Theorem:}}
The 1st-order error kets associated to the operators $\hat{A}+\hat{B}$, $c\hat{A}$ and $\hat{A}\hat{B}$ are respectively:
\begin{eqnarray}
|E(\hat{A}+\hat{B},\psi,a+b)> &=& |E(\hat{A},\psi,a)> + |E(\hat{B},\psi,b)>, \\
|E(c\hat{A},\psi,ca)> &=& c|E(\hat{A},\psi,a)>, \\
|E(\hat{A}\hat{B},\psi,ab)> &=& a|E(\hat{B},\psi,b)>+b|E(\hat{A},\psi,a)>+
|E(\hat{B},\hat{A},\psi,b,a)>.
\end{eqnarray}
\underline{{\bf Proof:}}
For the sum of operators we have:
\begin{equation}
|E_{A+B}>=(\hat{A}+\hat{B}-(a+b))|\psi>=
(\hat{A}-a)|\psi>+(\hat{B}-b)|\psi>=|E_{A}>+|E_B> .
\end{equation}
The proof of the product by a scalar and of the product of two operators follows the same lines, this time using the relations $(c\hat{A}-ca)=
c(\hat{A}-a)$ and:
\begin{equation}
(\hat{A}\hat{B}-ab)=a(\hat{B}-b)+b(\hat{A}-a)+(\hat{A}-a)(\hat{B}-b),
\end{equation}
that when applied to the state $|\psi>$ provide the desired results.$_{\Box}$

Let us now extend these results to the case of several products and sums of fundamental operators.
Let $\hat{A}$ be a general hermitian operator displayed as a sum of multiple products of fundamental operators and let $A^0$ be the classical function that is functionally identical to $\hat{A}$:
\begin{eqnarray}
\hat{A}=\sum_{i=1}^n c_i \hat{B}_i=\sum_{i=1}^n c_i\prod_{j=1}^m \hat{x}_{ij}
\qquad,\qquad
A^0=\sum_{i=1}^n c_i B_i^0=\sum_{i=1}^n c_i\prod_{j=1}^m x_{ij},
\end{eqnarray}
where $\hat{x}_{ij}$ is one of the fundamental operators
($\hat{x}_{ij}\in \{\hat I,\hat{q}_1...\hat{q}_N,\hat{p}_1,...\hat{p}_N\}$ where
$N$ is the dimension of the classical system), $n$ and $m$ are arbitrary positive integers, $c_i$ are complex numbers and $\hat{B}_i$,
$B_i$ are multiple products of the fundamental observables and multiple products of the corresponding classical observables, respectively. The map from $\hat{A}$ to $A^0$ will be named {\it unquantization}.
Clearly, the procedure (22) is beset by order problems (i.e. if $\hat{A}$ is displayed in different orders we get different $A^0$). The consequences of this will be discussed later on.

The aim now is to expand $(\hat{A}-A^0)$ in terms of $(\hat{x}_{ij}-
x_{ij})$. The first step is to put:
$(\hat{A}-A^0)=\sum_{i=1}^n c_i(\hat{B}_i-B^0_i)$.
Using eq.(21) we get:
\begin{eqnarray}
& & \hat{B}_i-B^0_i  =  \hat{x}_{i1}\prod_{j=2}^m \hat{x}_{ij}-
x_{i1}\prod_{j=2}^m x_{ij} \nonumber \\
& = & x_{i1} \left(\prod_{j=2}^m \hat{x}_{ij}-
\prod_{j=2}^m x_{ij} \right) + \prod_{j=2}^m x_{ij} (\hat{x}_{i1}-x_{i1})
+  (\hat{x}_{i1}-x_{i1}) \left(\prod_{j=2}^m \hat{x}_{ij}-
\prod_{j=2}^m x_{ij} \right),
\end{eqnarray}
and using (21) once again to expand $ (\prod_{j=2}^m \hat{x}_{ij}-
\prod_{j=2}^m x_{ij} ) $ in terms of $(\hat{x}_{i2}-x_{i2})$ and
$ (\prod_{j=3}^m \hat{x}_{ij}-\prod_{j=3}^m x_{ij} )$ and so on, after
using the relation (21) $m$ times we will obtain:
\begin{equation}
\hat{B}_i-B^0_i = \sum_{L=1}^{+\infty}
 \sum_{j_1 < j_2 .. < j_L=1}^m \frac{\partial^L B^0_i}{\partial x_{ij_1}...
\partial x_{ij_L}}(\hat{x}_{ij_1}-x_{ij_1})...(\hat{x}_{ij_L}-x_{ij_L})
\end{equation}
where the sum in $L$ can be truncated at the $m$th term.
The next step is to multiply each term in $i$ by $c_i$ and after sum in $i$.  We get:
\begin{equation}
\hat{A}-A^0 =  \sum_{L=1}^{+\infty} \sum_{i=1}^n
 \sum_{j_1 < j_2 .. < j_L=1}^m \frac{\partial^L A^0}{\partial x_{ij_1}...
\partial x_{ij_L}}(\hat{x}_{ij_1}-x_{ij_1})...(\hat{x}_{ij_L}-x_{ij_L})
\end{equation}
Notice that this is just a "Taylor expansion" of an operator around the classical observable with the same functional form.
The sums in the expansions (24) and (25) are taken over $j_1 < j_2 ... < j_L$ to preserve the order in which the operators $\hat{x}_{ij}$ appear in $\hat{B}_i$ since in general $(\hat{x}_{ik}-x_{ik})(\hat{x}_{ij}-x_{ij})
\not=(\hat{x}_{ij}-x_{ij})(\hat{x}_{ik}-x_{ik})$.
Because of this the analysis of the expansion (25) is tricky. The order in which the classical variables appear in $A^0$ is relevant. For instance if $\hat{A}=\hat{p}\hat{q}-\hat{q}\hat{p}$ then $A^0=pq-qp$ and this form, and not $A^0=0$, should be the one used to calculate the partial derivatives in (25).

Let us now consider the operator $(\hat{A})_+ = \sum_{i=1}^n c_i (\Pi_{j=1}^m \hat{x}_{ij})_+$ obtained from a general hermitian operator $\hat{A}$ (22) by a
term by term symmetrization:
$(\hat{A})_+ = \sum_{i=1}^n c_i (\hat{B}_{i})_+$,
where $(\hat{B}_{i})_+$ is the completly symmetric operator obtained from $\hat{B}_{i}$.
In general $(\hat{A})_+ \not= \hat{A}$ but
the classical observables obtained from $\hat{A}$ and $(\hat{A})_+$, given by  (22), are identical, i.e. ${A}^0 = ({A})_+^0$.
For this operator the expansion (25) might be written:
\begin{equation}
(\hat{A})_+ -A^0 =
\sum_{L=1}^{+\infty} \frac{1}{L!} \sum_{i=1}^n
 \sum_{j_1,j_2 ,.., j_L=1}^m \frac{\partial^L A^0}{\partial x_{ij_1}...
\partial x_{ij_L}}(\hat{x}_{ij_1}-x_{ij_1})...(\hat{x}_{ij_L}-x_{ij_L})
\end{equation}
Our next step is to understand under which conditions, if any, is the expansion (26) a valid approximation to the expansion (25).
To do this the first step is to define the unquantization map more precisely:

\underline{{\bf Definition}} {\bf - Unquantization $V_0$}\\
Let ${\cal A}({\cal H})$ be the algebra of linear operators acting on the physical Hilbert space ${\cal H}$ and let ${\cal A}(T^{\ast}M)$ be the algebra of complex functions over the classical phase space $T^{\ast}M$.
$V_0$ is the map:
\begin{equation}
V_0:{\cal A}({\cal H})\longrightarrow{\cal A}(T^{\ast}M);\quad A^0=V_0(\hat{A}),
\end{equation}
that satisfies the following {\it requirements}: \\
a) The action of $V_0$ on a fundamental operator provides the corresponding classical fundamental observable: $V_0(\hat{q}_i)=q_i$ and $V_0(\hat{p}_i)=p_i$, $i=1..N$.
Moreover $V_0(\hat{I})=1$.\\
b) The action of $V_0$ on a general operator $\hat{A}$, displayed in an arbitrary order, is given by $V_0(\hat{A})=V_0(\hat{A}_R)$ where $\hat{A}_R=\hat{A}$ but displayed in an order in which i) $\hat{A}$ is the sum of an hermitian term with an anti-hermitian term and
ii) all the commutators present in $\hat{A}$ have been resolved
($\hat{A}_R$ does not contain antisymmetric terms).\\
If $\hat{A}$ is displayed in the required order, $\hat{A}_R$ then:\\
c) $V_0$ is linear, i.e. if $\hat{A}_R=b\hat{B}+c\hat{C}$ then
$V_0(\hat{A}_R)=bV_0(\hat{B})+cV_0(\hat{C})$; $b,c \in {\cal C}$.\\
d) It is valid the product rule: if $\hat{A}_R=..+\hat{B}\hat{C}+..$ then
$V_0(\hat{A}_R)=V_0(..)+V_0(\hat{B})V_0(\hat{C})+V_0(..)_{\Box}$

One should notice that the map $V_0$ is not equivalent to the procedure (22). However, for a general operator $\hat A$, all the classical observables $V_0(\hat A)$ can also be obtained using the procedure (22) (by displaying $\hat A$ in appropriate order). Therefore all the results that were valid for $A^0$ obtained in (22), are also valid for $A^0=V_0(\hat A)$.
Further properties of the map $V_0$ will be discussed in section 4.

Let us proceed with the analysis of expansions (25,26). We realize that if $\hat{A}$ is displayed in the required order $\hat{A}_R$ then the difference between $\hat{A}$ and $(\hat A)_+$ is, at the most, proportional to a factor of $\hbar^2$. Since $A^0=V_0(\hat A)$ is identical to $V_0((\hat A)_+)$ we conclude that if $A^0=V_0(\hat A)$ then the difference between the right hand sides of equations (25) and (26) is at the most proportional to a factor of $\hbar^2$ which, in the context of the results of this paper is of negligible magnitude. In conclusion, if $A^0=V_0(\hat A)$ then the right hand side of (26) is a valid approximation to $\hat A -A^0$.

Let us proceed: since $\hat{x}_{ij}$ - in the expansion (26) - is one of the $2N$ fundamental operators the former expansion can be cast in the form:
\begin{eqnarray}
\hat{A}-A^0 &=& \sum_{L=1}^{+\infty} \frac{1}{L!}
\sum_{j_1,...,j_L=1}^{2N} \frac{\partial^L A^0}{\partial a_{j_1}...
\partial a_{j_L}}(\hat{a}_{j_1}-a_{j_1})...(\hat{a}_{j_L}-a_{j_L})
\end{eqnarray}
Finally, if we apply this expansion to the quantum state $|\psi>$,
we get:
\begin{eqnarray}
|E(\hat A,\psi,A^0)>&=&
\sum_{L=1}^{+\infty} \frac{1}{L!}
\sum_{j_1,...,j_L=1}^{2N} \frac{\partial^L A^0}{\partial a_{j_1}...
\partial a_{j_L}} |E_{a_{j1},...,a_{jL}}>
\end{eqnarray}
The generalization of the former set of results to the case of the $m$th-order
error ket, $|E^m_A>=(\hat{A}-A^0)^m|\psi>$
can be obtained by exponentiating the expansion (28) to the $m$th power. Up to the lowest order we get:
\begin{eqnarray}
|E^m_A>=(\hat{A}-A^0)^m|\psi> & = &
\sum_{k_1,...,k_m=1}^{2N} \left( \prod_{i=1}^m \frac{\partial A^0}
{\partial a_{k_i}}\right) |E_{a_{k1},...,a_{km}}>,
\end{eqnarray}
which is our final result concerning the error ket of an operator $\hat{A}$ functional of the fundamental operators. Notice that the former results are valid in general, irrespectively of the specific functional form of the wave function $|\psi>$.

\subsubsection{Evolving $\Delta_{m}$}

We shall now concentrate on the case of a system with a classical limit initial data. The aim is to calculate, in the representation of $\hat{A}$ (22), the value of the $m$th-order spread of a wave function $|\psi> $ satisfying (5).
To do this the main point is to calculate the norm of a general error ket:
$<E_{x_{1},....x_{n}}|E_{x_{1},....x_{n}}>$, where ${x}_{1},...,{x}_{n}$
 is an arbitrary sequence of fundamental observables: $x_i \in \{q_1,..,q_N,p_1,..p_N\},i=1..n, n\in {\cal N}$.
The following theorem will do this:

\underline{{\bf Theorem}}\\
If $|\psi>$ is a classical limit initial data wave function then the
norm of the error ket \\
$|E(\hat x_{1},..,\hat x_{n},\psi,x^0_1,..,x^0_n)>$ satisfies the following relation:
\begin{equation}
<E_{x_{1},....x_{n}} |E_{x_{1},....x_{n}}> \le \delta^2_{1}....\delta^2_{n},
\end{equation}
where $(x_i^0,\delta_i), i=1..n$ is the classical initial data associated to the observable $x_i$.\\
\underline{{\bf Proof:}}
This result will be proved by induction: \\
i) For $n=1$ and $x_1$ an arbitrary fundamental observable eq.(5) immediately implies (result 3.1-c): $<E_{x_1}|E_{x_1}>\le
\delta_1^2$.\\
ii) For an arbitrary $n$ we use (10) and write:
\begin{equation}
<E_{x_{1},....x_{n}} |E_{x_{1},....x_{n}}>=\sum_{s,x_n}
<E_{x_{1},....x_{n-1}}|x_n,s><x_n,s|E_{x_{1},....x_{n-1}}>|(x_n-x^0_n)|^2,
\end{equation}
where $x_n$ are the eigenvalues of the operator $\hat{x}_n$,
with degeneracy index $s$. We want to show that for all eigenvalues
$x_n \notin I_n$, where $I_n=[x_n^0-\delta_n,x_n^0+\delta_n]$, we have:
\begin{equation}
<x_{n},s |E_{x_{1},....x_{n-1}}>= <x_n,s|\prod_{i=1}^{n-1}
(\hat{x}_i-x^0_i)|\psi>=0, \quad \forall_{x_n \notin I_n}.
\end{equation}
To prove this let us expand the wave function $|\psi>$ using a second set of eigenstates of $\hat{x}_n$:
\begin{equation}
<x_n,s|\prod_{i=1}^{n-1}(\hat{x}_i-x^0_i)|\psi> =
\sum_{x_n^{\prime}\in I_n,s^{\prime}}<x_n,s|\prod_{i=1}^{n-1}
(\hat{x}_i-x^0_i)|x^{\prime}_n,s^{\prime}><x^{\prime}_n,s^{\prime}|\psi>,
\end{equation}
where $x_n^{\prime}$ are eigenvalues of $\hat{x}_n$ with degeneracy index $s^{\prime}$. Notice that in the representation of $\hat{x}_n$ the wave function is completely confined to the interval $I_n$. To prove the result (33) is then sufficient to show that:
\begin{equation}
<x_n,s|\prod_{i=1}^{j}\hat{x}_i|x^{\prime}_n,s^{\prime}>=0,\quad
\forall_{x_n\notin I_n,x^{\prime}_n \in I_n} , \forall_{j<n}.
\end{equation}
Let us then prove the former identity:
\begin{eqnarray}
<x_n,s|\prod_{i=1}^{j}\hat{x}_i|x^{\prime}_n,s^{\prime}>
  & = & \frac{1}{x^{\prime}_n}<x_n,s|\prod_{i=1}^{j}\hat{x}_i\hat{x}_n|
x^{\prime}_n,s^{\prime}> \nonumber \\
& = & \frac{1}{x^{\prime}_n}<x_n,s|\hat{x}_n \prod_{i=1}^{j}\hat{x}_i
+[\prod_{i=1}^{j}\hat{x}_i,\hat{x}_n]|x^{\prime}_n,s^{\prime}>
\end{eqnarray}
and if $<x_n,s|[\Pi_{i=1}^{j}\hat{x}_i,\hat{x}_n]|x^{\prime}_n,s^{\prime}>=0$
then:
\begin{eqnarray}
<x_n,s|\prod_{i=1}^{j}\hat{x}_i|x^{\prime}_n,s^{\prime}>=
\frac{1}{x^{\prime}_n}<x_n,s|\hat{x}_n
\prod_{i=1}^{j}\hat{x}_i|x^{\prime}_n,s^{\prime}>
=\frac{x_n}{x^{\prime}_n}<x_n,s|\prod_{i=1}^{j}\hat{x}_i|x^{\prime}_n,s^{\prime}>.
\end{eqnarray}
Since $x_n\not=x_n^{\prime}$ eq.(37) immediately implies (35) and thus the result (33) is valid.
The problem is now reduced to prove that $<x_n,s|[\Pi_{i=1}^{j}\hat{x}_i,
\hat{x}_n]|x^{\prime}_n,s^{\prime}>=0$, which in turn, and using the same procedure,
will be reduced to prove that $<x_n,s|[[\Pi_{i=1}^{j}\hat{x}_i,\hat{x}_n],
\hat{x}_n]|x^{\prime}_n,s^{\prime}>=0$, and so on until we obtain at the most a $j$-commutator which will always have the value zero (notice that $\hat{x}_i$
are fundamental operators).
Inserting the result (33) into (32) we get:
\begin{eqnarray}
<E_{x_{1},....x_{n}} |E_{x_{1},....x_{n}}> & = & \sum_{s,x_n \in I_n}
<E_{x_{1},....x_{n-1}}|x_n,s><x_n,s|E_{x_{1},....x_{n-1}}>|(x_n-x^0_n)|^2
\nonumber\\
 &\le& <E_{x_{1},....x_{n-1}}|E_{x_{1},....x_{n-1}}>\delta_{n}^2,
\end{eqnarray}
which proves the theorem.$_{\Box}$

A straightforward corollary of the former result is the one obtained by using the Schwartz inequality:
\begin{equation}
|<E_{x_{1},..x_{n}} |E_{y_{1},..y_{m}}>|  \le   \delta_{x_1}.. \delta_{x_n} \delta_{y_1}..\delta_{y_m},
\end{equation}
where ${x}_{1},..,{x}_{n}$ and ${y}_{1},..,{y}_{m}$ are two
arbitrary sequences of fundamental observables.

Let us return to the calculation of the $m$-order spread of the wave function
$|\psi>$, satisfying (5),
in the representation of $\hat{A}$. The relevant calculation is that of
$<E^m_A|E^m_A>$. From eq.(28,39), we get:
\begin{eqnarray}
<E^m_A|E^m_A> =<\psi|(\hat A-A^0)^{2m}|\psi>
\le \left( \sum_{L=1}^{+\infty} \frac{1}{L!} \sum_{j_1,...,j_L=1}^{2N}
\left|\frac{\partial^L A^0}{\partial a_{j_1}...\partial a_{j_L}}\right|\delta_{j_1}...\delta_{j_L} \right)^{2m} = \delta_{A^0}^{2m},
\end{eqnarray}
a result that is valid for all $m\in {\cal N}$. The $m$-order spread then reads:
\begin{equation}
\Delta_m(\hat A,\psi,A^0,p)=\left(\frac{<E^m_A|E^m_A>}{1-p}\right)^{1/2m} \le
\frac{\delta_{A^0}}{(1-p)^{1/2m}},
\end{equation}
which, for all $p<1$ converges to the
classical error margin $\delta_{A^0}$ as $m\rightarrow \infty$.

From the results (3.1-d)) and
(41) we can draw the conclusion that the classical limit initial data
wave function $|\psi>$,
in the representation of an arbitrary observable $\hat{A}$, is completely confined to
an interval around $A^0$ with the range of the classical error margin of
$A^0$, the relation between $\hat{A}$ and $A^0$ being the one given by (27).

\section{Unquantization}

In the last section we proved that the wave function $|\psi>$ - satisfying (5) - in the representation of a given quantum operator $\hat{A}$, which in the end is to be identified with $\hat{A}(t)$ given in (6), is completely confined to an interval centered at the value of the classical observable $A^0$ and with the range of the classical error interval of $A^0$.
This implies that the output of a measurement of $\hat{A}$, performed with an experimental apparatus of any resolution, will certainly belong to the previous error interval, which is exactly the classical prediction for the output of a measurement of the classical observable $A^0$.

However these are not our final results, yet. The reason is straightforward:
Let $\hat{A}=\hat{A}(t)$, what remains to be proven is simply that
 $A^0=V_0(\hat A(t))=A(t)$, i.e. that the classical observable $A^0$, obtained from the quantum observable $\hat{A}(t)$ using the map $V_0$, coincides with the classical observable $A(t)$ obtained by evolving $A(0)=V_0(\hat A(0))$ using the classical theory.

Hence the relevant questions are: Is the map $V_0$ well defined? And will it map
a quantum observable $\hat{A}(t)$ to the classical observable $A(t)$ given by (7)?

Starting with the first question it is easy to see that the map $V_0$ is not
well defined. In general $\hat{A}$ can be displayed in several different
functional forms (all of them satisfying the order requirement), each of which will be mapped by $V_0$ to a {\it different} (however very similar)
classical observable. A simple example will elucidate this point: let $\hat{A}=1/2(
\hat{x}\hat{y}\hat{z}+\hat{z}\hat{y}\hat{x})$ with $\hat{x}=\hat{q}_1\hat{p}_2$,
$\hat{y}=\hat{p}_1$ and $\hat{z}=\hat{q}_1\hat{q}_2$ where $q_1,p_1,q_2,p_2$ are the canonical variables of a two dimensional system. Alternatively, $\hat{A}$ might be written as $\hat{A}=
1/4(\hat{x}\hat{y}\hat{z}+\hat{z}\hat{y}\hat{x}+\hat{x}\hat{z}\hat{y}+
\hat{y}\hat{z}\hat{x}) + 1/4[\hat{x},[\hat{y},\hat{z}]]$. The first form of $\hat{A}$ is mapped by $V_0$ to the classical observable $A^0=xyz$ while the second form is mapped to the observable $A^0=xyz - 1/4 \hbar^2 q_1$.
Moreover for each different classical observable obtained we will, in general,
also get a different associated error ket and error margin.
This does not mean that the former results (30,41) concerning
the error ket and the spread of the wave function are incorrect. These results
have been proved to be valid (up to a correction term proportional to $\hbar^2$) for all different orders we may choose for the operator
$\hat{A}$, and consequently for all different $A^0$ obtained from $\hat{A}$, providing the order requirement is satisfied.

Notice however that this ambiguity would be problematic if the difference between two different classical observables, obtained from a single quantum one, had meaningful values. However, one can easily realize that if $A^0_1$ and $A^0_2$ are two such observables (i.e. $A^0_1=V_0(\hat A)$ and also $A^0_2=V_0(\hat A)$) then $A^0_1-A^0_2$ is proportional to a factor of, at the most $\hbar^2$. An imprecision of this magnitude is not meaningful when compared to the errors associated to the classical observables. That is the predictions $A^0_1$ and $A^0_2$ are well within the error interval of each other $|A^0_1 - A^0_2| << \delta_{{A_1^0} {\rm or }{A_2^0}}$. Hence, the two predictions are consistent with each other and we conclude that
$A^0_1$ and $A^0_2$ are equally valid candidates for a classical description of the quantum observable $\hat{A}$.

This takes us to the second question, which now can be restated as: will the classical observable $A(t)$, given by (7), be one of the images of $V_0(\hat{A}(t))$?

To answer this question let us start by presenting a second proposal for the unquantization map:

\underline{{\bf Definition}} {\bf - Unquantization $V$}\\
Using the notation of the previous definition we define the new
unquantization $V$ to be the map:
\begin{equation}
V:{\cal A}({\cal H})\longrightarrow{\cal A}(T^{\ast}M);\quad A=V(\hat{A}),
\end{equation}
that satisfies the following requirement: $V\circ \wedge=1$ where $\wedge$ is the Dirac quantization map \cite{dirac1,dirac2}. That is $V$ is
the inverse
of the Dirac quantization map $\wedge$.
The properties of $V$ follow immediately from the properties of $\wedge$: \\
a) $V(\hat{q}_i)=q_i$, $V(\hat{p}_i)=p_i$, $i=1..N$, and $V(\hat{I})=1$.\\
b) $V(1/i\hbar[\hat{A},\hat{B}])=\{V_0(\hat{A}),V_0(\hat{B})\}$ for all
 $\hat{A}$ and $\hat{B}$.\\
For a general operator $\hat{A}$ displayed in the required order (see definition of $V_0$):\\
c) $V$ is linear: if $\hat{A}=b\hat{B}+\hat{C}$ then $V(\hat{A})=bV(\hat{B})+V(\hat{C})$,
$b \in {\cal C}$.\\
d) It is valid the product rule: if $\hat{A}=..+\hat{B}\hat{C}+..$ then
$V(\hat{A})=V(..)+V(\hat{B})V(\hat{C})+V(..)$.$_{\Box}$

We should point out that, since the Dirac quantization map
$\wedge$ is not injective, the unquantization map $V$ is also
non-univocous. The simple example above - $\hat{A}=1/2(
\hat{x}\hat{y}\hat{z}+\hat{z}\hat{y}\hat{x})$ - can also be used
here to make this point clear. Being beset by the same type of
order problems still $V$ displays an important advantage over
$V_0$: it is straightforward to recognize that when $V$ is applied
to the operator $\hat{A}(t)$ -given by (6)- yields the classical
observable $A(t)$:
\begin{equation}
V(\hat{A}(t))=V\left(\sum_{n=0}^{\infty}\frac{1}{n!}
 \left({\frac{t}{i\hbar}}\right)^n [...[\hat{A},\hat{H}],\hat{H}]...]\right)=
\sum_{n=0}^{\infty}\frac{t^n}{n!} \{...\{A,H\},H\}...\}.
\end{equation}
Let us now consider a general operator $\hat{X}$. We want to prove that $V(\hat{X})$ provides a set of classical observables that is included in the set $V_0(\hat{X})$. Clearly the action of the two maps on an operator that does not contain antisymmetric components is identical. Moreover
the two commutation relations $1/i\hbar[,]$ and $\{,\}$ have "compatible" algebraic structures, that is we can resolve $1/i\hbar [\hat{A},\hat{B}]$ and express the result in such an order that when we perform the substitution of the quantum observables by the corresponding classical ones (i.e. when we unquantize $1/i\hbar [\hat{A},\hat{B}]$ using the map $V_0$) we get exactly the same final result as if we just compute $\{A,B\}$ (i.e. as if we use the map $V$ to unquantize $1/i\hbar [\hat{A},\hat{B}]$).
Thus, if $\hat{X}$ contains antisymmetric components the classical observable $V(\hat{X})$ might be obtained by displaying $\hat{X}$ in an adequate order and calculating $V_0(\hat{X})$.
Hence, it is always possible to obtain the classical observables $V(\hat{X})$ using the map $V_0$. This is the result we were looking for. It means that expansion (28) and thus all subsequent results are valid for $A^0=V(\hat{A})$.

Finally, we are able to state that the quantum mechanical predictions for an arbitrary
dynamical system with a classical limit initial data wave function are that
a measurement of an arbitrary fundamental observable $\hat{A}(t)$ at the time $t$
will yield the value given by (43) with an error margin given by (41) (with $m \to \infty$).
These are exactly the predictions of the classical mechanical treatment of the
same system.

\section{Criteria of classicality}

In the last section we proved that a dynamical system with a quantum initial data satisfying
(5) will evolve {\it exactly} accordingly to the predictions
of classical mechanics, that is the system is $\infty$-consistent for all times.
The main interest of this result is formal since, in most cases we are unable to provide a wave function satisfying the classical limit criterion (5).

To proceed we will now study dynamical systems with general physical initial data.
The aim is to use the formalism developed in the previous sections to derive two criteria providing a measure of the degree of classicality of an arbitrary quantum system with arbitrary initial data.

Let then $|\phi>$ be the initial data wave function of a $N$ dimensional dynamical system with canonical variables $a_k,(k=1..2N)$.
Let $\hat{S}_{k_i}=(\hat{a}_{k_1}, \hat{a}_{k_2}, .......\hat{a}_{k_n})$, $k_i \in \{1..2N\}$ be
a sequence of fundamental operators associated to the $n$-terms sequence
$k_i$ ($i=1..n;n \in {\cal N}$ ).
The relevant quantities that we have to calculate will be the $n$-order mixed error kets associated to the sequences $\hat{S}_{k_i}$:
\begin{equation}
|E_{S_{ki}}>=(\hat{a}_{k_1}-a^0_{k_1}) (\hat{a}_{k_2}-a^0_{k_2})....
 (\hat{a}_{k_n}-a^0_{k_n})|\phi> ,
\end{equation}
where $a^0_{k_i}$ is the classical initial value of the canonical variable $a_{k_i}$, i.e. $a^0_{k_i}=a_{k_i}(t=0)$.

\subsection{First Criterion}

The first step is to obtain the time evolution of the canonical variables using the standard classical formulation of the system:
\begin{equation}
a_j(t)=\sum_{m=0}^{\infty} \frac{t^m}{m!} \{....\{a_j,H\}....,H\} =
F_j(a_k,t).
\end{equation}
We then consider the sequences $S_{k_i}$ such that:
\begin{equation}
\frac{\partial F_j}{\partial S_{k_i}} = \frac{\partial^n F_j}{\partial a_{k_1}
\partial a_{k_2}.....\partial a_{k_n}} \not= 0,
\end{equation}
for at least one $j=1..2N$. For each of these sequences we construct the associated error ket (44).
This way we obtain a set of error kets. The first order classicality criterion is given by the following conditions:
\begin{equation}
<E_{S_{ki}}|E_{S_{ki}}> \le (\delta_{S_{ki}})^2 = (\delta_{k_1}
\delta_{k_2}.....\delta_{k_n})^2,
\end{equation}
where the inequalities should hold for all the sequences determined
in (46). In (47) $\delta_{k_i}$ is the initial data classical error margin associated to the classical observable $a_{k_i}$, $i=1..n$.
If the initial data wave function satisfies the former conditions
we say that the dynamical system is {\it first order classical}.

To proceed we construct the sequences $S^M_{k_i}$ formed by $M \in {\cal N}$
original sequences $S_{k_i}$, that is $S^M_{k_i}
=S_{k_i},S_{k^{\prime}_i},....S_{k^{\prime \prime}_i}$,
where each of the $M$ former sequences $S_{k_i}$ might be any of the ones determined in (46).
Once again, we calculate the values of
$<E_{S^M_{ki}}|E_{S^M_{ki}}> $ and compare them with
$(\delta_{S^M_{ki}})^2$. If the initial data satisfies:
\begin{equation}
<E_{S^M_{ki}}|E_{S^M_{ki}}>  \le (\delta_{S^M_{ki}})^2,
\end{equation}
for all possible sequences $S^M_{k_i}$ we say that the dynamical system is $M$-order classical.

Let us make two remarks: the first one to say that the classification of a given initial data as $M$-order classical is {\it dependent} on the scales (error margins $\delta_k$) that characterize the classical description. In particular for $ \delta_k = \infty $ all dynamical systems will be $\infty$-order classical (and will have a classical limit initial data) while for scales smaller than the Planck scale none will even be $1$st-order classical. The second remark is to point out that if a system is $M$-order classical then (result 3.1-d) its initial data wave function, in the representation of any of the observables $\hat{a}_k(0)$,
will have a minimum probability $p$ in the interval $I_k=[a_k(0)-\delta _k /
(1-p)^{1/(2M)} , a_k(0)+\delta _k / (1-p)^{1/(2M)} ]$, where
$a_k(0)$ is the value of the classical observable $a_k$ at
the initial time, $t=0$. That is the initial data
is $M$-order consistent. This is because in (46) we always determine the $2N$ single value sequences $S_{k_1}=a_k$, ($k=1..2N$).

Furthermore, if the classical and the quantum initial data satisfy the inequalities (48) then we can substitute $<E_{S^M_{ki}}|E_{S^M_{ki}}>$ by $ (\delta_{S^M_{ki}})^2$
when computing (40) for $A^0=a_j(t)$. The result (41) can then be easily obtained being valid up to the order $m=M$. Hence, using the result 3.1-d) we can state that an arbitrary $M$-order classical system,
as defined above, will evolve in such a way that in the representation
of $\hat{a}_j(t)$ (for all $j=1..2N$ and for all $t$)
the initial data wave function has at least a probability
$p$ confined to the interval $I_j(t)=[a_j(t)-\delta_j(t) /(1-p)^{1/(2M)},a_j(t)+\delta_j(t) /(1-p)^{1/(2M)}]$
with $a_j(t)$ and $\delta_j(t)$ being the classical time evolution of
the canonical variable $a_j$ and associated error margin.
According to the second consistency criterion this means that the classical and the quantum data describing the system at the time $t$ are $M$-order consistent.

We conclude that if $M$ is the order of classicality of a given dynamical system then: firstly the classical and the quantum data describing the initial time configuration of the system are $M$-order consistent and, secondly the degree of consistency is preserved through the time evolution.

Finally, notice that the higher the order of classicality $M$ of the dynamical system the more similar will be the range of $I_j(t)$ and the classical error margin.
When $M$ goes to infinity we obtain the classical limit description of the system.

\subsection{Second Criterion}

A second classicality criterion can be easily devised.
Once again the first step is to calculate the time evolution of the canonical variables $a_j$ using the classical formulation of the theory (45). Again we use this result to obtain the sequences of fundamental operators $\hat{S}_{k_i}$ (46) and the associated error kets (44).
Using these error kets we can write the second classicality criterion:
\begin{equation}
<E_{S_{ki}}|E_{S_{ki}}> \le (\delta_{S_{ki}})^2(1-p_0) = (\delta_{k_1}
\delta_{k_2}.....\delta_{k_n})^2 (1-p_0), \quad \forall S_{k_i} \quad \mbox{in (46)}.
\end{equation}
If the classical and the quantum initial data satisfy the former inequalities for a given $p_0$ ($0\le p_0 < 1$) and for all sequences determined in (46), then we say that the dynamical system is $p_0$-order classical.
A straightforward use of the result (3.1-d) will prove that a $p_0$-order
classical system is described by an initial data wave function
that, in the representation of any of its fundamental observables $\hat{a}_k(t=0), k=1..2N$, has at least a probability $p_0$ confined to the classical error interval
$[a_k(0)-\delta_k,a_k(0)+\delta_k]$.
That is the initial time configuration is $p_0$-order consistent (according to the first consistency criterion).

Now, let us concentrate on the dynamical evolution of the former initial data.
If the inequalities (49) are satisfied we can perform the substitution of $<E_{S_{ki}}|
E_{S_{ki}}>$ by $(\delta_{S_{ki}})^2(1-p_0)$ when computing (40) for $m=1$. Notice that there will be an extra factor of $(1-p_0)$ in the final expression in (40).
We can then proceed and obtain the result (41) for
$\Delta_1(\hat{a}_j(t),\phi , a_j(t),p)$ (which still contains the extra factor $(1-p_0)^{1/2}$).
More precisely we make $p=p_0$ in (41) and get for all $j=1..2N$:
\begin{eqnarray}
&& \Delta_1(\hat{a}_j(t),\phi , a_j(t),p_0) = \left(\frac{<E_{a_j(t)}|E_{a_j(t)}>}{1-p_0}\right)^{1/2} \le \nonumber \\
&& \le  \frac{1}{(1-p)^{1/2}} \sum_{L=1}^{+\infty} \frac{1}{L!} \sum_{k_1,...,k_L=1}^{2N}
\left|\frac{\partial^L a_j(t)}{\partial a_{k_1}...\partial a_{k_L}}\right|
\delta_{k_1}...\delta_{k_L}(1-p_0)^{1/2} =  \delta_{j}(t).
\end{eqnarray}

The previous result together with 3.1-d) implies that in the representation of $\hat{a}_j(t)$ (for all $j=1..2N$) the initial data wave function has at least a probability $p_0$ confined to the classical error interval $[a_j(t)-\delta_{j}(t),a_j(t)+\delta_{j}(t)]$, i.e. the classical and the quantum descriptions of the configuration of the system at the time $t$ are $p_0$-order consistent. This result is valid for all times. We conclude that if a dynamical system is $p_0$-order classical then the classical and the quantum descriptions of the configuration of the system are $p_0$-order consistent for all times.
Once again, when $p_0 \to 1$ we obtain the classical limit description of the system.

\section{Example - Harmonic Oscillator}

To illustrate the use of the first classicality criterion (section 5.1), let us obtain the classicality conditions
for the simple example of the harmonic oscillator. The classical Hamiltonian is
given by: $H=\frac{1}{2} (q^2+p^2)$,
where $q$ and $p$ are a pair of canonical variables and, to make it simple we made $w=m=1$.
Solving the equations of motion we obtain the classical time evolution of the canonical variables and the corresponding error margins:
\begin{equation}
\left\{ \begin{array}{l}
q (t) =  q(0) \cos t + p(0) \sin t \\
p(t) =  q(0) \sin t + p(0) \cos t
 \end{array} \right. \quad , \quad
\left\{ \begin{array}{l}
\delta_q(t) =  |\cos t| \delta_q(0) + |\sin t| \delta_p(0) \\
\delta_p(t) =  |\sin t| \delta_q(0) + |\cos t| \delta_p(0)
\end{array} \right.
\end{equation}
Let $|\phi>$ be the initial data wave function for the quantum harmonic oscillator.
Let us then determine the conditions that $|\phi>$ should satisfy so that the quantum system allows for a consistent $M$-order classical description. Following the general prescription of section 5.1 the first step is to determine the fundamental sequences (46). They are the single value sequences:
\begin{equation}
S_1=q \quad {\rm and} \quad S_2 = p.
\end{equation}
For a $M$-order classical system the relevant sequences are arrays of $M$ fundamental sequences:
\begin{equation}
S^{M}=(z_1,...,z_M), \quad z_i = q \vee p, \quad i=1..M,
\end{equation}
and the condition of $M$-order classicality (48) reads:
\begin{eqnarray}
&& <E_{S^{M}}|E_{S^{M}}> \le \delta^2_{S^{M}} \quad , \forall S^{M} \mbox{in (53)}\\
&& \Longleftrightarrow
<\phi|(\hat{z}_1-z_1(0))...(\hat{z}_M-z_M(0))(\hat{z}_M-z_M(0))...(\hat{z}_1-z_1(0))
|\phi> \le \delta_{z_1}^2(0)... \delta_{z_M}^2(0) \nonumber .
\end{eqnarray}
Given the classical initial data $\{q(0),p(0),\delta_q(0),\delta_p(0)\}$, eq.(54) constitute a system of inequalities to be satisfied by initial data wave function $|\phi>$.
For a first-order classical system the classicality conditions take their simpler form:
\begin{equation}
\left\{ \begin{array}{l}
<\phi|(\hat{q}-q(0))^2|\phi>  \le  \delta^2_q(0) \\
<\phi|(\hat{p}-p(0))^2|\phi>  \le  \delta^2_p(0)
\end{array} \right.
\Longleftrightarrow
\left\{ \begin{array}{l}
\int (q-q(0))^2 |\phi (q)|^2 dq  \le  \delta_q^2(0) \\
\int (p-p(0))^2 |\phi (p)|^2 dp  \le  \delta^2_p(0)
\end{array} \right.
\end{equation}
To obtain explicit solutions we may want to consider Gaussian wave packets (notice however that, in general, there are many solutions of (55) which are not Gaussians):
\begin{equation}
\phi_c(q_0,p_0,\Delta q, q)= \frac{1}{(2\pi (\Delta q)^2)^{1/4}} \exp \left\{
-\frac{(q-q_0)^2}{4 (\Delta q)^2} + i p_0 q /\hbar \right\},
\end{equation}
where $q_0,p_0$ and $\Delta q$ are parameters and the wave function was displayed in the $\hat q$ representation. If we make $q_0=q(0)$ and $p_0=p(0)$
and substitute $\phi_c(q(0),p(0),\Delta q,q)$ in (55) we get the equivalent set of inequalities:
\begin{equation}
\Delta q \le \delta_q(0) \quad \wedge \quad
\frac{\hbar}{2^{1/2} \Delta q} \le \delta_p(0).
\end{equation}
As expected, if $\delta_p(0) \delta_q(0) < \hbar/(2^{1/2})$ there is no coherent state (and actually no wave function) that might satisfy the former inequalities, while for larger values of the classical error margins there are many solutions of (55), including the coherent states with a parameter $\Delta q$ satisfying (57). We see that the classicality criteria provide a comparative notion of classicality but not an absolute one. In fact the degree of classicality is always relative to the classical description supplied.

The main result of the formalism is that a wave function satisfying (55), and in particular a coherent state satisfying (57) displays a time evolution first order consistent with the classical predictions (51), which means that for all $t$:
\begin{equation}
\int^{q(t)+\frac{\delta_q(t)}{(1-P)^{1/2}}}_{q(t)-\frac{\delta_q(t)}{(1-P)^{1/2}}}
|\phi(q,t)|^2 dq \ge P , \quad \wedge \quad
\int^{p(t)+\frac{\delta_p(t)}{(1-P)^{1/2}}}_{p(t)-\frac{\delta_p(t)}{(1-P)^{1/2}}}
|\phi(p,t)|^2 dp \ge P , \quad \forall_{0\le P <1},
\end{equation}
where $q(t),p(t), \delta_q(t),\delta_p(t)$ are given by (51), $|\phi(t)>$ is the solution of the Schr\"{o}dinger equation:
$$
i\hbar \partial /\partial t |\phi(t)> = 1/2(\hat{q}^2+\hat{p}^2)|\phi(t)>,
\quad  |\phi(t=0)>=|\phi>,
$$
and $P$ is an arbitrary probability. Take for instance $P=0.99$, eq.(58) states that
$99\%$ of the probability of the wave function $|\phi(t)>$, in both the representations of $\hat q$ and $\hat p$, is confined to the classical intervals $[q(t)-10\delta_q(t),q(t)+10\delta_q(t)]$ and $[p(t)-10\delta_p(t),p(t)+10\delta_p(t)]$, respectively. This statement is valid for all times and, for the case of coherent states can be easily verified by numerical computation of the integrals (58).

To see what happens when the degree of classicality is increased let us consider a 10th-order classical system. In this case the initial data wave function should satisfy the conditions (54) for all $M=10$ sequences (53). If we consider Gaussian type solutions (56) and substitute (56) in (54) we get the following set of inequalities for the parameter $\Delta q$:
\begin{equation}
\frac{(2M-1)!}{2^{M-1}((M-1)!)}(\Delta q)^{2M} \le \delta^{2M}_q(0) \quad \wedge \quad
\frac{(2M-1)!}{2^{M-1}((M-1)!)}\left(\frac{\hbar}{2^{1/2} \Delta q}\right)^{2M} \le \delta^{2M}_p(0),
\end{equation}
where $M=10$. Clearly, for the same classical initial data the set of solutions of (59) is just a subset of the set of solutions of (57).

The solutions of (54) with $M=10$ and in particular the coherent states satisfying (59), display a time evolution $10$th-order consistent with the classical predictions (section 5.1). In particular for $P=0.99$ we have $(1-P)^{1/20}=0.79$ and thus:
\begin{equation}
\int^{q(t)+1.25\delta_q(t)}_{q(t)-1.25\delta_q(t)}
|\phi(q,t)|^2 dq \ge 0.99  \quad \wedge \quad
\int^{p(t)+1.25\delta_p(t)}_{p(t)-1.25\delta_p(t)}
|\phi(p,t)|^2 dp \ge 0.99 ,
\end{equation}
where $|\phi(t)>$ is the quantum time evolution of a general initial data wave function satisfying (54) for $M=10$. The result is valid for all times and can be checked explicitly for an arbitrary coherent state satisfying (59).

One of the most interesting properties of coherent states is that its quantum time evolution is (in some sense) "consistent" with the classical predictions. Because of this coherent states might be seen as "classical states".
We see that the classicality criteria provide a systematic procedure to obtain "classical states" for an arbitrary dynamical system: by solving the classicality conditions (54) we obtain a large set of classical states (much larger than the coherent states set) for which the quantum time evolution is consistent with the classical predictions. Moreover, and most important, the imprecise notions of "consistency" and of "classical states" are made fully precise in this formalism.

\section{Further example}

To further illustrate the use of the criteria let
us consider the 2-dimensional system described by the Hamiltonian:
\begin{equation}
H=\frac{P^2}{2M} + \frac{p^2}{2m} +kQp^2,
\end{equation}
where $(Q,P)$ are the canonical variables of a particle of mass $M$, $(q,p)$ the ones of the particle of mass $m$ and $k$ is a coupling constant.
The classical time evolution is obtained by solving the Hamiltonian equations of motion which yield:
\begin{equation}
\left\{ \begin{array}{l}
Q(t)=Q(0)+\frac{P(0)}{M}t-\frac{k}{2M}p(0)^2t^2 \\
P(t)=P(0)-kp(0)^2t \\
q(t)=q(0)+\{ \frac{p(0)}{m}+2kQ(0)p(0)\} t+\frac{k}{M}P(0)p(0)t^2 - \frac{k^2}{3M}p(0)^3t^3 \\
p(t)=p(0)
\end{array}  \right.
\end{equation}
with error margins (taking into account all orders in the initial time error margins):
\begin{equation}
\left\{ \begin{array}{lll}
\delta_Q(t)& = & \delta_Q(0)+\left|\frac{t}{M} \right| \delta_P(0) + \left|\frac{kp(0)t^2}{M} \right| \delta_p(0) +\left|\frac{kt^2}{2M} \right| \delta^2_p(0)\\
\delta_P(t) & = &\delta_P(0)+\left|{2kp(0)t} \right| \delta_p(0) +\left|{kt} \right| \delta^2_p(0) \\
\delta_q(t) & = &\delta_q(0) +\left|(\frac{1}{m}+2kQ(0))t+\frac{kP(0)t^2}{M}-\frac{k^2p(0)^2t^3}{M} \right| \delta_p(0)+\left|2kp(0)t \right|\delta_Q(0)+\\
&+& \left|\frac{kp(0)t^2}{M} \right|\delta_P(0)+
|kt|\delta_Q(0) \delta_p(0) +\left|\frac{kt^2}{2M} \right| \delta_P(0)\delta_p(0) +
\left|\frac{k^2p(0)t^3}{M}\right| \delta_p^2(0) + \left|\frac{k^2t^3}{M} \right| \delta^3_p(0) \\
\delta_p(t) & = & \delta_p(0)
\end{array} \right.
\end{equation}
For this dynamical system the fundamental sequences (46) are (the indexes $1,2,3$ and $4$ refer to the canonical variables $q,p,Q$ and $P$, respectively):
\begin{eqnarray}
&& S_1=q \quad , \quad
S_2=p \quad , \quad
S_3=Q \quad , \quad
S_4=P \quad , \quad \\
&& S_{22}=(p,p) \quad , \quad
S_{23}=(p,Q) \quad , \quad
S_{24}=(p,P) \quad , \quad
S_{222}=(p,p,p) \nonumber
\end{eqnarray}
and the first order classicality condition reads:
\begin{equation}
<E_{S_{ki}}|E_{S_{ki}}> \le \delta^2_{S_{ki}} \quad , \quad \forall S_{k_i} \quad {\rm in \quad (64)}
\end{equation}
To make the discussion simpler let us consider solutions of the form $|\xi>=|\psi>|\phi>$, where $|\psi>$ is the quantum state of the particle of mass $m$, and $|\phi>$ is that of the particle of mass $M$. In the $(\hat q, \hat Q)$ representation we have $\xi (q,Q)=\psi(q)\phi(Q)$. To proceed we
notice that if the condition (65) is satisfied for the sequences $S_1,S_2,S_3$ and $S_4$ then is also satisfied for $S_{23}$ and $S_{24}$. Furthermore, if it is satisfied for $S_{222}$ is also for $S_{22}$ and $S_{2}$. Hence, the system of eight inequalities (65) is reduced to the system of four inequalities: $
<E_{S_{ki}}|E_{S_{ki}}> \le \delta^2_{S_{ki}} \quad , \quad S_{ki} \in \{S_1 ,S_3 ,S_4, S_{222} \} $. Moreover, the former system splits into two independent systems:
\begin{equation}
\left\{ \begin{array}{l}
 <\psi|(\hat{q}-q(0))^2|\psi> \le \delta^2_q(0) \\
 <\psi|(\hat{p}-p(0))^6|\psi> \le \delta^6_p(0)
\end{array} \right. \quad {\rm and } \quad
\left\{ \begin{array}{l}
 <\phi|(\hat{Q}-Q(0))^2|\phi> \le \delta^2_Q(0) \\
 <\phi|(\hat{P}-P(0))^2|\phi> \le \delta^2_P(0)
\end{array}  \right.
\end{equation}
For Gaussian type solutions $\xi_c(Q_0,P_0,q_0,p_0,\Delta Q,\Delta q,q,Q)=
\phi_c(Q_0,P_0,\Delta Q,Q)\psi_c(q_0,p_0,\Delta q,q)$
it reduces to the form:
\begin{equation}
\left\{ \begin{array}{l}
\Delta q \le \delta_q(0) \\
\left( 15 \right)^{1/6} \frac{\hbar}{2^{1/2} \Delta q} \le \delta_p(0)
\end{array} \right. \quad {\rm and } \quad
\left\{ \begin{array}{l}
\Delta Q \le \delta_Q(0) \\
\frac{\hbar}{2^{1/2} \Delta Q} \le \delta_P(0)
\end{array}  \right.
\end{equation}
where we used eq.(59) to obtain the second inequality of the first system. Any Gaussian wave function satisfying the former conditions displays a time evolution
(solution of the Schr\"{o}dinger equation: $i\hbar \partial / \partial t |\xi(t)> = \hat{H} |\xi (t)>$, $|\xi(0)>=|\xi>$) first order consistent with the classical predictions (62,63). This means that for any of the canonical variables $Z=q,p,Q \vee P$ and for all times:
\begin{equation}
\int_{w} \int^{Z(t)+\delta_Z(t)/(1-P)^{1/2}}_{Z(t)-\delta_Z(t)/(1-P)^{1/2}}
|<\xi(t)|z,w>|^2 dz dw \ge P, \quad \forall_{0 \le P <1}
\end{equation}
where $|z,w>$ is the general eigenstate of the observable $\hat Z$ with associated eigenvalue $z$ and degeneracy index $w$, $P$ is a probability and ($Z(t),\delta_Z(t)$) are the classical time evolution of the canonical variable $Z$ and its error margin which are given by eqs.(62,63), respectively.

Just like in the example of the harmonic oscillator we can increase the order of classicality of the system by requiring the initial data wave function to satisfy higher order classicality conditions. Likewise, this will increase the degree of consistency between the classical and the quantum predictions.

\section{Conclusions}

A general procedure leading to the construction of a measure of classicality was presented. The procedure can be summarized in two main steps:\\
1) Definition of a consistency criterion. This is a kinematical criterion. It should provide a suitable measure of the degree of agreement between the classical and the quantum descriptions of a single, specific time configuration of the system. Two intuitive definitions of consistency criteria were presented and certainly other, possible more interesting criteria might be defined.\\
2) Using the consistency criterion the problem of studying the overall consistency between the classical and the quantum descriptions of a general dynamical system is reduced to a simpler and more precisely formulated problem. That of identifying the conditions that should be satisfied so that the degree of consistency between the classical and the quantum initial data is preserved through time evolution.
These conditions were explicitly obtained and used to construct a classicality criterion for each of the consistency criteria proposed at the beginning. We expect that the same general procedure might be used to derive the classicality criterion associated to other proposals of consistency criteria.

Ultimately, the two measures of classicality obtained are arbitrary in nature since
they are a direct consequence of the definition of a consistency criterion.
Therefore, they should be tested in physical examples to see if they yield sensible results.
The criteria of this paper were applied to two simple dynamical systems and yield expected results. Namely, Gaussian wave functions were obtained as "classical-like states" (solutions of the classicality conditions).

Further applications of the criteria have been developed in \cite{nuno2,nosso1,nosso2}. Other possible future applications which might be of considerable interest are in the field of non-linear dynamical systems, where the connection between classical and quantum mechanics becomes more involved.

To finish let us make a few remarks: \\
i) The degree of classicality of a given quantum system is always relative to the classical description supplied (more precisely, to the classical initial data supplied). Therefore, the criteria do not provide an absolute classification of classicality but only a comparative one.\\
ii) The fully classical behavior (i.e. the complete consistency with the classical predictions) is obtained in the limit case in which the quantum system fully satisfies one of the classicality criteria to all orders or equivalently, if the quantum and the classical initial data are fully consistent. Notice that consistency and classicality, in this case and in general only in this case, are equivalent notions: the results of section 3.2.2 imply this statement directly.\\
iii) The classicality conditions vary from one dynamical system to another. More precisely, once the classical initial data is given the $M$-order classicality conditions on the quantum initial data are more restrictive for some systems (for instance, the second example) than for others (for instance, the first example). This is not surprising, as neither is the fact that for some dynamical systems, the classicality conditions are so restrictive that there is no wave function that, for typical values of the classical data, might satisfy even the first order classicality conditions (consider for instance the system of a particle in a potential step). \\
iv) The classicality criteria provide only a partial answer to the problem of measuring the degree of classicality of a general dynamical system. We were able to prove that a general dynamical system will evolve in agreement with the classical predictions up to some
degree if its initial time configuration satisfies the classicality
conditions up to some extent. However, these are sufficient but not necessary conditions: the system may very well not satisfy even the first order classicality criterion and still present a fairly classical-like evolution.

\subsection*{Acknowledgments}

I would like to thank Jorge Pullin for many suggestions
and insights made through the present work.
I would also like to thank Gordon Fleming, Don Marolf, Jo\~{a}o Prata, Troy Schilling and Lee Smolin for several discussions and insights in the subject.

This work was supported by funds provided by Junta Nacional de
Investiga\c{c}\~{a}o Cient\'{\i}fica e Tecnol\'{o}gica -- Lisbon -- Portugal,
grant B.D./2691/93 and by grants NSF-PHY 94-06269,
NSF-PHY-93-96246, the Eberly Research fund at Penn State and
the Alfred P. Sloan foundation.

\end{document}